\documentclass[10pt,floatfix]{revtex4}
                                                                                
\usepackage{graphicx}
\usepackage{mathrsfs}
\usepackage{amsmath}
\usepackage{amsfonts}
\usepackage{amssymb}
\usepackage{epsfig}                                                                                
\usepackage{theorem}
\theorembodyfont{\upshape}

\newcommand{\be}{\begin{equation}}
\newcommand{\ee}{\end{equation}}
\newcommand{\bea}{\begin{eqnarray}}
\newcommand{\eea}{\end{eqnarray}}
\newcommand{\nn}{\nonumber \\}

\input epsf.sty

\newlength{\diaght}
\newlength{\diaghtwo}
\newlength{\diaghtthree}
\newlength{\diagshift}
\newcommand{\qed}{\nobreak \ifvmode \relax \else
      \ifdim\lastskip<1.5em \hskip-\lastskip
      \hskip1.5em plus0em minus0.5em \fi \nobreak
      \vrule height0.75em width0.5em depth0.25em\fi}

\begin{document}


\title{Form factor expansion of the row and diagonal correlation functions of the two dimensional Ising model}

\author{I. Lyberg
\footnote{e-mail ilyberg@math.sunysb.edu}
and B. M. McCoy 
\footnote{e-mail mccoy@max2.physics.sunysb.edu}}
\affiliation{$^{*}$Department of Mathematics, 
State University of New York, Stony Brook,  NY 11794-3840 \\
$^{\dagger}$Institute for Theoretical Physics, State University of New York,
 Stony Brook,  NY 11794-3840}
\date{\today}
\preprint{YITPSB-06-57}

\begin{abstract}
We derive and prove exponential and form factor expansions of the row correlation function ${\langle}\sigma_{0,0}\sigma_{0,N}{\rangle}$ and the diagonal correlation function ${\langle}\sigma_{0,0}\sigma_{N,N}{\rangle}$ of the two dimensional Ising model.
\end{abstract}

\maketitle                                                                             
                                                                               
\begin{flushright}
 {\tt YITP-SB-06-57}
\end{flushright}
                                                                                

\section{Introduction}
\label{intro}
The correlation functions ${\langle}\sigma_{0,0}\sigma_{M,N}{\rangle}$ 
of the two dimensional
Ising model with horizontal (vertical) interaction energies $E_1$
($E_2$) can be written in many different ways which appear to be
different but  which in fact are equal. They were first expressed as
determinants by Kaufman and Onsager \cite{KO}. Later Montroll, Potts and
Ward \cite{MPW} demonstrated that if an arbitrary path is drawn 
on the lattice connecting the point $(0,0)$ with 
the point $(M,N)$ then the correlation
can be expressed as a determinant whose size in general is twice the length of
the path. The correlations ${\langle}\sigma_{0,0}\sigma_{0,N}{\rangle}$ 
and ${\langle}\sigma_{0,0}\sigma_{N,N}{\rangle}$ can
both be expressed as $N\times N$ Toeplitz determinants 
\cite{KO}--\cite{ste}, and expressions of 
$\langle \sigma_{0,0}\sigma_{M,N}\rangle$ as
determinants of size $M$ and $M+1$ for $M\geq N$ were 
given by Yamada \cite{Ya2}, \cite{Ya3}. Furthermore
the correlations $\langle \sigma_{0,0}\sigma_{M,N}\rangle$ for 
all finite $M,~N$ 
were expressed as determinants of Fredholm operators 
by Cheng and Wu \cite{CW}. 

The representations of the correlations as finite size determinants
gives an efficient evaluation when the separation is small but to
investigate the large separation behavior alternative representations
are needed. The first such result is the limiting behavior for 
$T<T_c$
\bea
S_{\infty}=&&\lim_{N\rightarrow \infty}\langle \sigma_{00}\sigma_{0N} \rangle
=\lim_{N\rightarrow \infty}\langle \sigma_{00}\sigma_{NN}\rangle \nn
=&&\left\{1-(\sinh 2E_1/kT\sinh 2E_2/kT)^{-2}\right\}^{1/4},
\label{lb}
\eea
which is most easily computed \cite{MPW} by the use of Szeg{\"o}'s
  theorem \cite{SG},\cite{SG2}. 

The first large separation expansion for both $T<T_c$ and $T>T_c$ 
beyond the limiting value (\ref{lb}) was given in 1966 by Wu \cite{W}  
for $\langle \sigma_{00}\sigma_{0N}\rangle$
by applying a Wiener-Hopf procedure to the $N \times N$ Toeplitz
determinant representation. Shortly thereafter Cheng and Wu \cite{CW}
obtained the leading term of the large separation 
behavior of $\langle \sigma_{00}\sigma_{MN} \rangle$ by applying a Wiener-Hopf
procedure to the Fredholm determinant representation. This derivation
is formally valid only for $M\neq 0$, and even though it is expected
that the result of \cite{CW} with $M$ formally set equal 
to zero should agree with the 
result of \cite{W}, there is no analytic derivation in the literature
that for $T<T_c$ the two results are in fact equal (even though the
equality has been verified to large orders in the low temperature
expansion.)

The expansions of \cite{W} and \cite{CW} may be considered as the
first terms in a systematic expansion of the correlations. The expansion
technique of \cite{CW} which starts from the Fredholm determinant
representation was carried out to all orders by Wu, McCoy, Tracy and
Barouch \cite{M2} in 1976 where it is found that the correlations can
be written in the following exponential
representation 
 \begin{equation}
{\langle
  \sigma_{00}\sigma_{MN}\rangle}_{T<T_c}
=S_{\infty}\exp\sum_{n=1}^{\infty}F_{MN}^{(2n)}  ~~~ \textrm{for}~ T<T_c
\label{m1}
\end{equation}
and as
\be  \langle \sigma_{00}\sigma_{MN}\rangle_{T>T_c}
=\widehat{S}_{\infty}\sum_{m=0}^{\infty}G_{MN}^{(2m+1)}
\exp{\sum_{n=1}^{\infty}\widehat{F}^{(2n)}_{MN}}   ~~~ \textrm{for}~ T>T_c
\label{m2}
\ee
where
\be \widehat{S}_{\infty}
=
\left\{1-(\sinh{2E_1/kT}\sinh{2E_2/kT})^{2}\right\}^{1/4}.
\label{la}
\ee
In \cite{M2} the expressions for $F^{(j)}_{MN},~ {\widehat
  F}^{(j)}_{MN}$ and $G_{MN}^{(j)}$ are given as $2j$ fold multiple
  dimensional integrals.

The exponentials in (\ref{m1}) 
and (\ref{m2}) may be expanded to give what is called a form factor expansion
\be \langle \sigma_{00}\sigma_{MN}\rangle_{T<T_c}
=S_{\infty}\sum_{n=0}^{\infty}f^{(2n)}_{MN}~~~{\rm for}~~T<T_c
\label{m3}
\ee
and
\be  \langle \sigma_{00}\sigma_{MN}\rangle_{T>T_c}
=\widehat{S}_{\infty}\sum_{n=0}^{\infty}f^{(2n+1)}_{MN}~~~{\rm for}~~T>T_c.
\label{m4}
\ee
The first few terms in this expansion were given in \cite{M2}. In the
scaling limit $N\rightarrow \infty,~~T\rightarrow T_c$ with $N|T-T_c|$
fixed the full expansion was given by Nappi \cite{Na}. For fixed $N$
and $T<T_c$ the full expansion (\ref{m3})  was given by Palmer and Tracy
\cite{PT}. Both of the cases $T<T_c$ and $T>T_c$ were treated by Nickel 
\cite{Ni1} \cite{Ni2}. An independent expansion was given by 
Yamada \cite{Ya1}, and this is shown in
\cite{Ni2} to agree with the results from the expansion of the
exponential forms of \cite{M2}.

The results for the exponential representation of the correlations 
\cite{M2} were obtained by extending to all orders the interative
expansion of the Fredholm determinant representation
\cite{CW}. However, as noted above, the result of \cite{CW} 
for $F^{(2)}_{M,N}$   when specialized to $M=0$ ``looks different''
from the corresponding result for ${\langle
  \sigma_{0,0}\sigma_{0,N}\rangle}$ obtained in \cite{W}. Moreover the
leading order large $N$ behavior of 
${\langle} \sigma_{0,0}\sigma_{N,N}{\rangle}$ is  obtained \cite{mwbook} 
from the results for of \cite{W} for
${\langle}\sigma_{0,0}\sigma_{0,N}{\rangle}$
and this result looks very different from the result of
\cite{M2}. Therefore it must be the case that if the Wiener-Hopf
procedure of Wu \cite{W}, which starts from the $N\times N$
Toeplitz determinant representation of  
${\langle \sigma_{0,0}\sigma_{0,N}\rangle}$
and ${\langle \sigma_{0,0}\sigma_{N,N}\rangle},$ is  iterated to all
orders we will obtain a representation of 
${\langle \sigma_{0,0}\sigma_{0,N}\rangle}$
and ${\langle \sigma_{0,0}\sigma_{N,N}\rangle}$ 
which is different from that of \cite{M2}. The purpose of this paper
is to derive these exponential expansions for both $T<T_c$
and $T>T_c$ and the form factor expansions which follow from them.
The results for the case of the diagonal correlation have been 
previously presented by one of us (BM) in \cite{M}
with Boukraa, Hassani, Maillard, Orrick and Zenine.
This paper is the proof, derivation and generalization of those results.

In sec. \ref{sum} we summarize the results of our calculations. 
In section \ref{below} we calculate the exponential representation of
the correlation  
functions $\langle \sigma_{00}\sigma_{0N}\rangle$ and 
$\langle \sigma_{00}\sigma_{NN}\rangle$ for $T<T_c.$ In section 
\ref{above} we calculate the exponential representations for $T>T_c.$ 
In section \ref{below2} 
we calculate the form factor expansions of
$\langle \sigma_{00}\sigma_{0N}\rangle$ and 
$\langle \sigma_{00}\sigma_{NN}\rangle$ for $T<T_c$ and  
section \ref{above2} we calculate the 
form factor expansions for $T>T_c,$ We conclude in
sec. \ref{dis} with a brief discussion of our results.

\section{Summary of Results}
\label{sum}

 We let $D_{N}$ stand 
for $S_N=\langle \sigma_{00}\sigma_{0N}\rangle$ or 
$C_N=\langle \sigma_{00}\sigma_{NN}\rangle$. Then
\begin{eqnarray}
 D_N= \left\lbrace \begin{array}{ll}
            D_{N}^{(-)}  & \mbox{for $T<T_c$,}\\ 
            D_{N}^{(+)}  & \mbox{for $T>T_c$,} 
\end{array} \right.
\label{def}
\end{eqnarray}
The representation of these correlations as an $N\times N$ Toeplitz
determinant is  \cite{mwbook} 
\bea D_N=\det{\mathbf{A}_{N}}
\label{det}
\eea
where
\bea
\mathbf{A}_{N}=\begin{pmatrix}
a_{0} &  a_{-1}  & \ldots & a_{1-N}\\
a_{1}  &  a_{0} & \ldots & a_{2-N}\\
\vdots & \vdots & \ddots & \vdots\\
a_{N-1}  &   a_{N-2}       &\ldots & a_{0}
\end{pmatrix}
\eea
and 
\bea a_n=\frac{1}{2\pi i}\oint_{|z|=1} \varphi(z)z^{-n-1}~dz,
\eea
where the path of integration is counterclockwise. 
The function $\varphi(z)$ is 
\bea \varphi(z)=\left(\frac{(1-\alpha_1z)(1-\alpha_2z^{-1})}
{(1-\alpha_1z^{-1})(1-\alpha_2z)}\right)^{1/2}.
\label{gen}
\eea 
For the diagonal correlation function 
$C_N$
\bea \alpha_1=0~~~{\rm and}~~ \alpha_2=(\sinh{2K_1}\sinh{2K_2})^{-1}
\label{ac}
\eea
where $K_j=E_j/kT.$  For the row correlation 
function $S_N$
\bea \alpha_1=e^{-2K_2}\tanh{K_1} ~~{\rm and}~~\alpha_2=e^{-2K_2}\coth{K_1}.
\label{row}
\eea

When $T<T_c$, then $\alpha_1<\alpha_2<1$. 
In this case we write $\varphi$ in a factored form as
\bea \varphi (z)=P(z)^{-1}Q(z^{-1})^{-1}
\eea
where 
\bea 
P(z)=\left((1-\alpha_2z)/(1-\alpha_1z)\right)^{1/2}
\label{P}
\eea 
and
\bea Q(z)=\left((1-\alpha_1z)/(1-\alpha_2z)\right)^{1/2}=1/P(z)
\label{Q}
\eea
We will prove in Sec. \ref{below} that the correlation 
function $D_{N}^{(-)}$ has an 
exponential expansion 
\bea D_{N}^{(-)}=S_{\infty}\exp{\sum_{n=1}^{\infty}F^{(2n)}_N}
\label{effei}
\eea
where
\begin{equation}
S_{\infty}=\left[\frac{(1-\alpha_1^2)(1-\alpha_2^2)}
    {(1-\alpha_1\alpha_2)^2}\right]^{1/4}
\label{sinfb}
\end{equation}
which for both the diagonal (\ref{ac}) and row (\ref{row}) specializes to
(\ref{lb}), and
\bea F^{(2n)}_N=\frac{(-1)^{n+1}}{n(2\pi)^{2n}}
\lim_{\epsilon \rightarrow 0}\prod_{i=1}^{2n}\oint_{|z_i|=1-\epsilon}dz_i 
\prod_{j=1}^{2n}\frac{z_j^{N}}{1-z_jz_{j+1}}
\prod_{k=1}^n P(z_{2k})P(z_{2k}^{-1})Q(z_{2k-1})Q(z_{2k-1}^{-1})
\label{F}
\eea
and $z_{2n+1}=z_1$. This agrees with the result 
given in ref. \cite{M} for the diagonal correlation 
function $C_{N}^{(-)}$.

In Sec. \ref{below2} we prove that
$D_{N}^{(-)}$ has the form factor expansion 
\bea D_{N}^{(-)}=S_{\infty}\sum_{n=0}^{\infty}f^{(2n)}_N
\label{ffm}
\eea 
where $f^{(0)}_N=1$ and 
\bea 
f^{(2n)}_N=&&\frac{1}{(n!)^2(2\pi)^{2n}}
\lim_{\epsilon \rightarrow 0}\prod_{i=1}^{2n}\oint_{|z_i|=1-\epsilon}dz_i 
~z_i^{N}
\prod_{k=1}^n P(z_{2k})P(z_{2k}^{-1})Q(z_{2k-1})Q(z_{2k-1}^{-1})\nn
&&\prod_{l=1}^n\prod_{m=1}^n(1-z_{2l-1}z_{2m})^{-2}
\prod_{1\leq p<q\leq n}(z_{2p-1}-z_{2q-1})^2(z_{2p}-z_{2q})^2.
\label{f}
\eea
This agrees with the result given in ref. \cite{M} for the diagonal correlation function $C_{N}^{(-)}$.

For $T>T_c$, we consider a new function $\varphi_1(z)$ such that
\bea \varphi_1(z)=\varphi(z)z
=\left(\frac{(1-\alpha_1z)(1-\alpha_2^{-1}z)}
{(1-\alpha_1z^{-1})(1-\alpha_2^{-1}z^{-1})}\right)^{1/2}
\label{varphi1}
\eea
which we write in factored form as 
\bea \varphi_1(z)=\widehat{P}(z)^{-1}\widehat{Q}(z^{-1})^{-1}
\eea
with
\bea \widehat{P}(z)=((1-\alpha_1z)(1-\alpha_2^{-1}z))^{-1/2}
\label{Ph}
\eea 
and
\bea \widehat{Q}(z)=((1-\alpha_1z)(1-\alpha_2^{-1}z))^{1/2}=1/\widehat{P}(z).
\label{Qh}
\eea
$\widehat{P}(z)$ and $\widehat{Q}(z)$ are analytic and non-zero for $|z|<1$.

We prove in Sec. \ref{above} that
the correlation function $D_{N}^{(+)}$ has an exponential expansion
\be D_{N}^{(+)}=-{\widehat S}_{\infty}\sum_{m=0}^{\infty}G_N^{(2m+1)}
\exp{\sum_{n=1}^{\infty}\widehat{F}^{(2n)}_{N+1}}
\label{effeai}
\ee
where  
\begin{equation}
{\widehat
  S}_{\infty}
=\left[(1-\alpha_1^2)(1-\alpha_2^{-2})(1-\alpha_1\alpha_2^{-1})^2\right]^{1/4}
\label{sinfa}
\end{equation}
which for both the diagonal (\ref{ac}) and row (\ref{row})
correlations specializes to (\ref{la})
and
$\widehat{F}^{(2n)}_{N}$ is defined as in (\ref{F}), 
but with $P$ and $Q$ replaced by $\widehat{P}$ and $\widehat{Q}$. 
Thus we find from (\ref{F}) that $\widehat{F}^{(2n)}_N$ is 
\bea \widehat{F}^{(2n)}_N
=\frac{(-1)^{n+1}}{n(2\pi)^{2n}}\lim_{\epsilon \rightarrow 0}
\prod_{i=1}^{2n}\oint_{|z_i|=1-\epsilon}dz_i 
\prod_{j=1}^{2n}\frac{z_j^{N}}{1-z_jz_{j+1}}\prod_{k=1}^{n}
\widehat{P}(z_{2k})\widehat{P}(z_{2k}^{-1})\widehat{Q}(z_{2k-1})
\widehat{Q}(z_{2k-1}^{-1})
\label{Fh}
\eea
and $G^{(2n+1)}_N$ is given by
\bea G^{(2n+1)}_N=\frac{1}{(2\pi i)^{2n+1}}&&\lim_{\epsilon \rightarrow 0}\prod_{i=1}^{2n+1}\oint_{|z_i|=1-\epsilon} dz_i~z_i^{N+1}\frac{1}{z_1z_{2n+1}}\prod_{k=1}^{2n}\frac{1}{1-z_kz_{k+1}}\nn
&&\prod_{l=1}^{n+1}\widehat{P}(z_{2l-1})\widehat{P}(z_{2l-1}^{-1})\prod_{m=1}^{n}\widehat{Q}(z_{2m})\widehat{Q}(z_{2m}^{-1}).
\label{G}
\eea

Equations (\ref{Fh}) and (\ref{G}) agree with the results given in ref. \cite{M}.
Note that for the diagonal correlation function 
$C_{N}^{(+)}=\langle \sigma_{00}\sigma_{NN}\rangle$ (\ref{ac}) implies that 
\be \widehat{F}^{(2n)}_N=F^{(2n)}_N.
\label{c}
\ee

In Sec. \ref{above2} we prove  that $D_{N}^{(+)}$ 
has the form factor expansion 
\bea D_{N}^{(+)}=-{\widehat  S}_{\infty}\sum_{n=0}^{\infty}f^{(2n+1)}_N
\label{rffeai}
\eea 
where
\bea f_N^{(2n+1)}=-\frac{i}{n!(n+1)!(2\pi)^{2n+1}}\lim_{\epsilon \rightarrow 0}\prod_{i=1}^{2n+1}\oint_{|z_i|=1-\epsilon} dz_i~z_i^{N}\prod_{l=1}^{n+1}\widehat{P}(z_{2l-1})\widehat{P}(z_{2l-1}^{-1})z_{2l-1}^{-1}\prod_{m=1}^{n}\widehat{Q}(z_{2m})\widehat{Q}(z_{2m}^{-1})z_{2m}\nn
\prod_{p=1}^{n+1}\prod_{q=1}^{n}\frac{1}{(1-z_{2p-1}z_{2q})^2}\prod_{1\leq j<k\leq n+1}(z_{2j-1}-z_{2k-1})^2\prod_{1\leq r<s\leq n}(z_{2r}-z_{2s})^2.
\label{fa6n}
\eea
Equation (\ref{fa6n}) agrees with 
result given in ref. \cite{M} for the diagonal correlation 
function $C_{N}^{(+)}$.


The proofs of these results are not restricted to the Ising case where
the generating function is given by (\ref{gen}) but with a suitable
replacement for the factors $S_{\infty}$ and ${\widehat S}_{\infty}$  
are valid in 
more general cases, for example the XY model in a magnetic field 
\cite{lsm}-\cite{bar}. The results (\ref{effei})-(\ref{f}) for $T<T_c$
are valid for any generating function $\varphi(z)$ where $\log \varphi(z)$ is
analytic and periodic on $|z|=1$ and $P(z)=1/Q(z)$
The results (\ref{effeai})-(\ref{G}) and (\ref{effeai})-(\ref{fa6n}) 
for $T>T_c$ are similarly valid for any generating function
for which $\log z\varphi(z)$ is analytic and periodic on the unit circle $|z|=1$ and 
${\widehat P}(z)=1/{\widehat Q}(z).$

\section{The exponential expansion for $T<T_c$}
\label{below}

In this section, we will use the theory of Wiener-Hopf 
sum equations to prove that the functions $F^{(2n)}_N$ which 
appear in equation (\ref{effei}) are given by (\ref{F}).

When $T<T_c$, then $\alpha_1<\alpha_2<1$ and therefore $P(z)$ and $Q(z)$ are analytic and non-zero for $|z|<1$. Furthermore the index of $\varphi$ is
\bea {\rm Ind}\varphi= \log{\varphi(e^{2\pi i})}-\log{\varphi(1)}=0
\label{index0}
\eea
It follows from (\ref{index0}) that we may use Szeg\"{o}'s 
theorem to find
\begin{equation}
\lim_{N \rightarrow \infty}D_{N}^{(-)}=S_{\infty} 
\end{equation}
with $S_{\infty}$ given (\ref{sinfb}) which reduces to (\ref{lb})
for both the diagonal and the row correlation functions. 
Therefore we may write
\bea D_{N}^{(-)}=S_{\infty}\prod_{n=N}^{\infty}D_{n}^{(-)}/D_{n+1}^{(-)}
\label{prodx}
\eea

\subsection{Computation of the ratio $D_{N}^{(-)}/D_{N+1}^{(-)}$}
\label{ratios}

The ratio $D_{N}^{(-)}/D_{N+1}^{(-)}$ is given by
\bea D_{N}^{(-)}/D_{N+1}^{(-)}=x_0^{(N)}
\label{xoratio}
\eea 
where $\mathbf{x}^{(N)}=(x_0,x_1,...,x_{N})$ satisfies
\bea \mathbf{A}_{N+1}\mathbf{x}^{(N)}=\mathbf{d}^{(N)}
\label{vectorA}
\eea
and $d_i^{(N)}=\delta_{i0}$. We indicate that 
the vector $\mathbf{x}^{(N)}$ has $N+1$ entries by writing $x_0^{(N)}$.

We will calculate $x_{0}^{(N)}$ by iterating the 
procedure given by Wu in section 3 of reference \cite{W}. 

\vspace{.1in}

{\bf Lemma 1} There are functions $\phi_N^{(2n)}$ such that
\bea x_{0}^{(N)}=1+\sum_{n=1}^{\infty}\phi_N^{(2n)}
\label{x0N}
\eea
 where
\bea 
\phi_N^{(2n)}=\frac{(-1)^{n+1}}{(2\pi)^{2n}}
\lim_{\epsilon \rightarrow 0}\prod_{i=1}^{2n}
\oint_{|z_i|=1-\epsilon}dz_i ~z_i^{N+1}\frac{1}{z_1z_{2n}}
\prod_{k=1}^n Q(z_{2k-1})Q(z_{2k-1}^{-1})P(z_{2k})
P(z_{2k}^{-1})\prod_{l=1}^{2n-1}\frac{1}{1-z_{l}z_{l+1}}.
\label{phi2n}
\eea

{\bf Proof} Let $h(\xi)$ be a function defined on the unit circle 
$|\xi|=1$, and let $h(\xi)$ have the Laurent expansion

\bea h(\xi)=\sum_{n=-\infty}^{\infty}h_n\xi^n.
\eea
From this we define 



\bea [h(\xi)]_+=\sum_{n=0}^{\infty}h_n\xi^n,~~~~~
[h(\xi)]_-=\sum_{n=-\infty}^{-1}h_n\xi^n,~~~~~\textrm{and}~~~~
[h(\xi)]_+'=\sum_{n=1}^{\infty}h_n\xi^n.
\label{hp}
\eea
From equations (\ref{hp}) it follows that 
\bea [h(\xi^{-1})]_-=[h(\xi)]_+'.
\label{hpm}
\eea

Equations (\ref{hp}) have the integral representations
\bea [h(\xi)]_+=\frac{1}{2\pi i}\lim_{\epsilon \rightarrow 0}\oint_{|\xi'|=1+\epsilon}~d\xi'\frac{h(\xi')}{\xi'-\xi},
\eea
\bea [h(\xi)]_-=\frac{1}{2\pi i}\lim_{\epsilon \rightarrow 0}\oint_{|\xi'|=1-\epsilon}~d\xi'\frac{h(\xi')}{\xi-\xi'},
\eea
and
\bea [h(\xi)]_+'=&&[h(\xi)]_+-\frac{1}{2\pi i}\oint_{|\xi|=1} d\xi~\frac{h(\xi)}{\xi}\nn
=&&\frac{1}{2\pi i}~ \xi \lim_{\epsilon \rightarrow 0}\oint_{|\xi'|=1+\epsilon} d\xi'~\frac{h(\xi')}{\xi'(\xi'-\xi)}.
\label{pb}
\eea
We define
\begin{equation}
X_N(\xi)=\sum_{n=0}^{N-1}x_n^{(N)}\xi^n
\end{equation}
It has been proven by Wu \cite{W} that the ratio (\ref{xoratio}) is given by 
\bea x_{0}^{(N)}=X_N(0)
\label{X0}
\eea
where $X_N(\xi)$ is a function determined by equations 
(2.19a)-(2.20b) of reference \cite{W} (with $Y(\xi)=1$). These equations are
\bea X_N(\xi)=P(\xi)\left\{[Q(\xi^{-1})]_++[Q(\xi^{-1})U_N(\xi)\xi^N]_+\right\} 
~~~~~(2.19\textrm{a}),
\label{X}
\eea

\bea V_N(\xi^{-1})=-(Q(\xi^{-1}))^{-1}\left\{[Q(\xi^{-1})]_-
+[Q(\xi^{-1})U_N(\xi)\xi^N]_-\right\} ~~~~~(2.20\textrm{a}), 
\label{V}
\eea

\bea X_N(\xi^{-1})\xi^N=Q(\xi)\left\{[P(\xi^{-1})\xi^N]_+
+[P(\xi^{-1})V(\xi)\xi^N]_+\right\}~~~~~~(2.19\textrm{b}),
\label{Xm}
\eea
and

\bea U_N(\xi^{-1})=-(P(\xi^{-1}))^{-1}\left\{[P(\xi^{-1})\xi^N]_-
+[P(\xi^{-1})V_N(\xi)\xi^N]_-\right\}~~~~(2.20\textrm{b}).
\label{U}
\eea

For our purposes we use equations (\ref{Q}), (\ref{hpm}) and the equality
$[Q(\xi^{-1})]_+=1$ to rewrite equations (\ref{X}), (\ref{V}) and (\ref{U}) as
\bea X_N(\xi)=P(\xi)\left\{1+[Q(\xi^{-1})U_N(\xi)\xi^N]_+\right\},
\label{Xo}
\eea

\bea V_N(\xi^{-1})=-P(\xi^{-1})\left\{[Q(\xi^{-1})]_-+[Q(\xi^{-1})U_N(\xi)\xi^N]_-\right\}
\label{Vo},
\eea
and
\bea U_N(\xi)=-Q(\xi)\left\{[P(\xi)\xi^{-N}]_+'+[P(\xi)V_N(\xi^{-1})\xi^{-N}]_+'\right\}\label{Ui}
\eea
We define $V_N^{(1)}(\xi^{-1})$ by replacing $U_N(\xi)$ by 0 in equation (\ref{Vo}). Thus
\bea V_N^{(1)}(\xi^{-1})=-P(\xi^{-1})[Q(\xi^{-1})]_-.
\label{V1}
\eea
We note from equation (\ref{Q}) that $Q(0)=1$. Thus because $Q(\xi^{-1})$ is analytic for $|\xi|>1$, we have 
\bea [Q(\xi^{-1})]_-=Q(\xi^{-1})-Q(0)=Q(\xi^{-1})-1.
\label{Qeq}
\eea
Therefore it follows from equations (\ref{Q}) and (\ref{Qeq}) that 
\bea -P(\xi^{-1})[Q(\xi^{-1})]_-=P(\xi^{-1})-1,
\label{Peq}
\eea
and thus equation (\ref{V1}) becomes
\bea V^{(1)}(\xi^{-1})=P(\xi^{-1})-1.
\label{V1pn}
\eea 
We define $U^{(1)}(\xi)$ by replacing $V_N(\xi^{-1})$ in (\ref{Ui}) $V_N^{(1)}(\xi^{-1})$ as given by equation (\ref{V1pn}). Thus  we find 
\bea U_N^{(1)}(\xi)=-Q(\xi)[P(\xi^{-1})P(\xi)\xi^{-N}]_+'
\label{U1p}
\eea
It follows from equation (\ref{Xo}) that $X_N^{(1)}(\xi)$ is given by

\bea X_N^{(1)}(\xi)&&=P(\xi)\Big\{1-\big[Q(\xi^{-1})Q(\xi)[P(\xi^{-1})P(\xi)\xi^{-N}]_+'\xi^N\big]_+\Big\}\nn
&&=P(\xi)\left\{1-\frac{1}{2\pi i}\lim_{\epsilon \rightarrow 0}\oint_{|\xi'|=1+\epsilon} d\xi'\frac{\xi'^{N}}{\xi'-\xi}Q(\xi^{-1})Q(\xi)[P(\xi^{-1})P(\xi)\xi^{-N}]_+'\right\}.
\label{X1}
\eea
Letting $\xi=0$ in equation (\ref{X1}), and using $P(0)=1$, and writing $X^{(1)}(0)=1+\phi_N^{(2)}$ we obtain

\bea \phi_N^{(2)}&&=-\frac{1}{2\pi i}\oint_{|\xi|=1} d\xi~Q(\xi^{-1})
Q(\xi)[P(\xi^{-1})P(\xi)\xi^{-N}]_+'\xi^{N-1}\nn
&&=-\frac{1}{2\pi i}\lim_{\epsilon \rightarrow 0}\oint_{|\xi|=1} d\xi_1~Q(\xi_1^{-1})Q(\xi_1)\frac{1}{2\pi
  i}\xi_1^N
\oint_{|\xi_2|=1+\epsilon} d\xi_2\frac{1}{\xi_2}\frac{1}{\xi_2-\xi_1}
P(\xi_2^{-1})P(\xi_2)\xi_2^{-N}.
\label{X1z}
\eea
Thus, if we set 
\bea \xi_{2k+1}=z_{2k+1},~~~~\xi_{2k}=z^{-1}_{2k}
\label{cov}
\eea
we obtain $\phi_N^{(2)}$ as given by equation (\ref{phi2n}).

We now calculate $V_N^{(2)}(\xi^{-1})$ by using 
equation (\ref{U1p}) in equation (\ref{Vo}): 
\bea V_N^{(2)}(\xi^{-1})&&=-P(\xi^{-1})
\left\{[Q(\xi^{-1})]_-+[Q(\xi^{-1})U_N^{(1)}(\xi)\xi^N]\right\}\nn
&&=-P(\xi^{-1})[Q(\xi^{-1})]_-
+P(\xi^{-1})\big[Q(\xi^{-1})Q(\xi)\xi^N[P(\xi^{-1})P(\xi)\xi^{-N}]_+'\big]_-.
\label{V2}
\eea

Next, we calculate $U_N^{(2)}(\xi)$ by using 
equation (\ref{V2}) in equation (\ref{Ui}): 
\bea U_N^{(2)}(\xi)&&=-(P(\xi))^{-1}\left\{[P(\xi)\xi^{-N}]_+'
+[P(\xi)V_N^{(2)}(\xi^{-1})\xi^{-N}]_+'\right\}\nn
&&=-Q(\xi)[P(\xi)P(\xi^{-1})\xi^{-N}]_+'
-Q(\xi)\Big[P(\xi)P(\xi^{-1})\xi^{-N}\big[Q(\xi)Q(\xi^{-1})\xi^N
[P(\xi)P(\xi^{-1})\xi^{-N}]_+'\big]_-\Big]_+'.
\label{U2ii}
\eea

We will now calculate $X_N^{(2)}(\xi)$ from (\ref{Xo}) as
\bea X_N^{(2)}(\xi)&&=P(\xi)\{1+[Q(\xi^{-1})U_N^{(2)}(\xi)\xi^N]_+\}\nn
&&=P(\xi)-P(\xi)\big[Q(\xi^{-1})Q(\xi)[P(\xi)P(\xi^{-1})\xi^{-N}]_+'\xi^N\big]_+\nn
&&-P(\xi)\bigg[Q(\xi^{-1})Q(\xi)\xi^N\Big[P(\xi)P(\xi^{-1})\xi^{-N}
\big[Q(\xi)Q(\xi^{-1})\xi^N[P(\xi)P(\xi^{-1})\xi^{-N}]_+'\big]_-\Big]_+'\bigg]_+.
\label{X2}
\eea

Letting $\xi=0$ in equation (\ref{X2}), 
we obtain $X_N^{(2)}(0)=1+\phi_N^{(2)}+\phi_N^{(4)}$:
\bea \phi_N^{(4)}   &&=-\frac{1}{2\pi i}\oint_{|\xi_1|=1} d\xi
~Q(\xi^{-1})Q(\xi)\Big[P(\xi^{-1})P(\xi)\xi^{-N}\big[Q(\xi^{-1})Q(\xi)\xi^{N}
[P(\xi^{-1})P(\xi)\xi^{-N}]_+'\big]_-\Big]_+'\xi^{N-1}\nn
&&=-\frac{1}{(2\pi i)^4}\lim_{\epsilon \rightarrow 0}
\oint_{|\xi|=1} d\xi_1~\xi_1^NQ(\xi_1^{-1})Q(\xi_1)
\oint_{|\xi_2|=1+\epsilon} d\xi_2~\frac{1}{\xi_2-\xi_1}\xi_2^{-N-1}
P(\xi_2^{-1})P(\xi_2)\nn
&&\oint_{|\xi_3|=1} d\xi_3~\frac{1}{\xi_3-\xi_2}\xi_3^{N+1}
Q(\xi_3^{-1})Q(\xi_3)\oint_{|\xi_4|=1+\epsilon} d\xi_4~\frac{1}{\xi_4-\xi_3}\xi_4^{-N-1}P(\xi_4^{-1})P(\xi_4).
\label{X2z}
\eea
Using the change of variables (\ref{cov}) 
we obtain an equation agreeing with equation (\ref{phi2n}).

In general, we iteratively define (from equation \ref{Vo})

\bea V_N^{(n+1)}(\xi^{-1}):
=-P(\xi^{-1})\left\{[Q(\xi^{-1})]_-+[Q(\xi^{-1})U_N^{(n)}(\xi)\xi^N]_-\right\}.
\label{Vn}
\eea
It then follows from equation (\ref{Ui}) that
\bea U_N^{(n)}(\xi)- U_N^{(n-1)}(\xi)
=-Q(\xi^{-1})\bigg[P(\xi)P(\xi^{-1})\xi^{-N}
\Big[Q(\xi)Q(\xi^{-1})\xi^N\big[P(\xi)P(\xi^{-1})\xi^{-N}
[Q(\xi)Q(\xi^{-1})\xi^N...]_-\big]_+'\Big]_-\bigg]_+',
\eea
where there are $2n-1$ brackets. It now follows from 
equations (\ref{X0}) and and (\ref{Xo}) that $\phi_N^{(2k)}$ is

\bea \phi_N^{(2k)}=-\frac{1}{2\pi i}
\oint_{|\xi|=1} d\xi~\xi^{N-1}Q(\xi)Q(\xi^{-1})\bigg[P(\xi)P(\xi^{-1})\xi^{-N}
\Big[Q(\xi)Q(\xi^{-1})\xi^N\nn
\big[P(\xi)P(\xi^{-1})\xi^{-N}[Q(\xi)Q(\xi^{-1})\xi^N...]_-\big]_+'\Big]_-\bigg]_+',
\eea
where there are $2k-1$ brackets. By use of (\ref{cov}), 
one obtains equation (\ref{phi2n}). This ends the proof of the lemma.

\subsection{Exponentiation}
\label{rep}

To complete the proof of the exponential form (\ref{effei}) we need to
  use (\ref{prodx}), (\ref{xoratio}) and (\ref{x0N}) to compute 
$F^{(2n)}_N$ as given in (\ref{F}).
We  begin by defining a function

\bea 
\tilde{F}^{(2n)}_N:=\frac{(-1)^{n+1}}{n(2\pi)^{2n}}\lim_{\epsilon \rightarrow 0}\prod_{i=1}^{2n}\oint_{|z_i|=1-\epsilon}dz_i 
\prod_{j=1}^{2n}\frac{z_j^{N}}{1-z_jz_{j+1}}\prod_{l=1}^{n}Q(z_{2l-1})Q(z_{2l-1}^{-1})P(z_{2l})P(z_{2l}^{-1})\left(1-\prod_{k=1}^{2n}z_{k}\right)
\eea
(We define $\tilde{F}_N^{(0)}=0$). Clearly
\bea 
F^{(2n)}_N=\sum_{k=N}^{\infty}\tilde{F}^{(2n)}_k.
\label{Ft}
\eea
Let $\phi_N^{(2n)}$ be given by equation (\ref{phi2n}) 
when $n\geq 1$ and let $\phi_N^{(0)}=1$. We define the functions 

\bea \phi(\lambda):=\sum_{n=0}^{\infty}\phi_{N}^{(2n)}\lambda^n
\label{phisum}
\eea
and
\bea {\tilde F}(\lambda):=\sum_{n=0}^{\infty}\tilde{F}_{N}^{(2n)}\lambda^n.
\label{Fsum}
\eea
Clearly $\phi(0)=1$ and $F(0)=0$. We would like to show that 

\bea \phi(\lambda)=e^{{\tilde F}(\lambda)}
\label{phiF}
\eea
It follows as a special case of (\ref{phiF}) with $\lambda=1$ that
\bea X_N(0)=\exp{\sum_{k=1}^{\infty}\tilde{F}_N^{(2k)}},
\eea 
and hence it follows from equations (\ref{prodx}) and (\ref{Ft}) that
\bea C_N&&=
(1-t)^{1/4}\exp{\sum_{k=N}^{\infty}\sum_{n=1}^{\infty}\tilde{F}_k^{(2n)}}
=(1-t)^{1/4}\exp{\sum_{n=1}^{\infty}\sum_{k=N}^{\infty}\tilde{F}_k^{(2n)}}
=(1-t)^{1/4}\exp{\sum_{n=1}^{\infty}F_N^{(2n)}}
\eea
This proves equation (\ref{F}). It remains to show that equation (\ref{phiF}) holds. 
Since $\phi(0)=1$ and $F(0)=0$, equation (\ref{phiF}) 
is equivalent to the equation
\bea \phi'(\lambda)=e^{{\tilde F}(x)}{\tilde F}'(\lambda)
\label{dphiF}
\eea
It follows from equations (\ref{phisum}), (\ref{Fsum}) and (\ref{dphiF}) that equation (\ref{phiF}) is equivalent to the following equation:

\vspace{.1in}

{\bf Lemma 2}
\bea 
n\phi_N^{(2n)}=\sum_{l=1}^nl\tilde{F}_N^{(2l)}\phi_N^{(2n-2l)}
\label{eq1}
\eea

{\bf Proof} It follows from (\ref{phi2n}) that the left hand side of (\ref{eq1}) is
\bea 
n\phi_N^{(2n)}=n(-1)^{n+1}\frac{1}{(2\pi)^{2n}}\lim_{\epsilon \rightarrow 0}\prod_{i=1}^{2n}\oint_{|z_i|=1-\epsilon} dz_i~z_i^{N+1}\frac{1}{z_{2n}z_1}\prod_{j=1}^{n}P(z_{2j})P(z_{2j}^{-1})Q(z_{2j-1})Q(z_{2j-1}^{-1})\prod_{k=1}^{2n-1}\frac{1}{1-z_kz_{k+1}},
\label{eq2l}
\eea 
and the right hand side is 
\bea \sum_{l=1}^nl\tilde{F}_N^{(2l)}\phi_N^{(2n-2l)}
&&=(-1)^n\frac{1}{(2\pi)^{2n}}\lim_{\epsilon \rightarrow 0}\prod_{i=1}^{2n}\oint_{|z_i|=1-\epsilon} dz_i\prod_{j=1}^{2n}
\frac{z_j^{N}}{1-z_jz_{j+1}}
\prod_{q=1}^{n}P(z_{2q})P(z_{2q}^{-1})Q(z_{2q-1})Q(z_{2q-1}^{-1})\nn
&&\left\{\sum_{l=1}^{n-1}\frac{1}{1-z_1z_{2l}}
\left(1-\prod_{k=1}^{2l}z_k\right)
(1-z_{2l}z_{2l+1})(1-z_{2n}z_1)
\prod_{m=2l+2}^{2n-1}z_{m}
-\left(1-\prod_{p=1}^{2n}z_p\right)\right\},
\label{eq2r}
\eea
where the product $\prod_{m=2l+2}^{2n-1}z_{m}$ is such that it equals
1 when $l=n-1$. Note that the product 
$\prod_{j=1}^{2n}$ is symmetric both in even and in odd variables separately.
 Hence $1-\prod_{k=1}^{2n}z_{k}$ can be rewritten  (under integration) as
\bea 1-\prod_{k=1}^{2n}z_{k}\equiv (1-z_1z_{2n})
\left(1+\sum_{q=1}^{n-1}\prod_{r=2}^{2q+1}z_r\right).
\eea
Next, note that the factor $\frac{1}{1-z_1z_{2l}}(1-z_{2l}z_{2l+1})(1-z_{2n}z_1)
\prod_{m=2l+2}^{2n-1}z_{m}$ does not involve any of the 
variables $\{z_i\}_{i=2}^{2l-1}$. Hence the product $1-\prod_{k=1}^{2l}z_{k}$ can be rewritten as
\bea 1-\prod_{k=1}^{2l}z_{k}\equiv (1-z_1z_{2l})
\left(1+\sum_{q=1}^{l-1}\prod_{r=2}^{2q+1}z_r\right).
\eea
Then the relevant factor of the integrand of the right hand 
side of equation (\ref{eq2r}) becomes
\bea 
&&(1-z_{2n}z_1)\left\{
\sum_{l=1}^{n-1}(1-z_{2l}z_{2l+1})\left(1+\sum_{q=1}^{l-1}\prod_{r=2}^{2q+1}z_r\right)
\prod_{m=2l+2}^{2n-1}z_{m}-\left(1+\sum_{q=1}^{n-1}\prod_{r=2}^{2q+1}z_r\right)\right\}      \nn
&&=(1-z_{2n}z_1)
\left\{\sum_{l=1}^{n-1}\left(1+\sum_{q=1}^{l-1}\prod_{r=2}^{2q+1}z_r\right)
\left(\prod_{m=2l+2}^{2n-1}z_{m}-\prod_{m=2l}^{2n-1}z_{m}\right)-\left(1+\sum_{q=1}^{n-1}\prod_{r=2}^{2q+1}z_r\right)\right\}.
\label{eq81}
\eea
After expansion of the first summand the right hand side of (\ref{eq81}) 
becomes
\bea
(1-z_{2n}z_1)
\left\{\sum_{l=1}^{n-1}\prod_{m=2l+2}^{2n-1}z_m-\sum_{l=1}^{n-1}\prod_{r=2}^{2n-1}z_r-\left(1+\sum_{q=1}^{n-1}\prod_{r=2}^{2q+1}z_r\right)\right\}
\label{eq82}
\eea
under integration. After summation (\ref{eq82}) becomes
\bea
-n(1-z_{2n}z_1)\prod_{r=2}^{2n-1}z_r,
\eea
which completes the proof. The proof of lemma 2 concludes the proof of equation (\ref{F}).

\section{The exponential expansion for $T>T_c$}
\label{above}

In this section, we will prove that the functions $\widehat{F}^{(2n)}_N$ and $G^{(2n+1)}_N$ in (\ref{effeai}) are given by (\ref{Fh}) and (\ref{G}). We will follow the procedure of section 2 of Wu \cite{W}. 
When $T>T_c$, then $\alpha_1<1<\alpha_2$ and $\varphi_1(z)$ has index 0. 
We define 
\bea b_n:=\frac{1}{2\pi i}\oint_{|z|=1} \varphi_1(z)z^{-n-1}~dz=a_{n-1}
\label{b}
\eea
We further define
\bea \mathbf{B}_{N+1}:=
\begin{pmatrix}
b_{0} &  b_{-1}  & \ldots & b_{-N}\\
b_{1}  &  b_{0} & \ldots & b_{1-N}\\
\vdots & \vdots & \ddots & \vdots\\
b_{N}  &   b_{N-1}       &\ldots & b_{0}
\end{pmatrix}
\label{mat}
\eea
and
\bea \widehat{D}_{N+1}:=\det{\mathbf{B}_{N+1}}
\label{deth}
\eea
We note that if we remove the first row and the last column from
$\widehat{D}_{N+1}$ and use (\ref{b}) we obtain $D_N$ as defined by
(\ref{det}). Therefore we may write

\bea D_{N}^{(+)}=\frac{D_{N}^{(+)}}{\widehat{D}_{N+1}}\widehat{D}_{N+1}=(-1)^Nx_{N}^{(N)}{\widehat D}_{N+1},
\eea 
where the ratio $D_{N}/\widehat{D}_{N+1}$ is given as

\bea \frac{D_{N}^{(+)}(t)}{\widehat{D}_{N+1}(t)}=(-1)^{N}x_N^{(N)}
\label{xNs}
\eea
and $\mathbf{x}^{(N)}=(x_0,x_1,...,x_N)$ satisfies
\bea \mathbf{B}_{N+1}\mathbf{x}^{(N)}=\mathbf{d}^{(N)}
\label{vector}
\eea
and $d_i^{(N)}=\delta_{i0}$. We indicate that the vector $\mathbf{x}^{(N)}$ has $N+1$ entries by writing $x_N^{(N)}$. Since $\varphi_1(z)$ has index 0, it follows from Szeg\"{o}'s theorem that 
\bea \lim_{N\rightarrow \infty}(-1)^N\widehat{D}_N={\widehat S}_{\infty}
\label{szegoe}
\eea
where ${\widehat S}_{\infty}$ is given by (\ref{sinfa}).
Thus, exactly as for $T<T_c$, 
\bea (-1)^{N+1}\widehat{D}_{N+1}={\widehat S}_{\infty}\prod_{n=N+1}^{\infty}\frac{\widehat{D}_{n}}{\widehat{D}_{n+1}}.
\eea
Furthermore the ratio $\widehat{D}_{n}/\widehat{D}_{n+1}$ and the product 
\bea \prod_{n=N+1}^{\infty}\frac{\widehat{D}_{n}}{\widehat{D}_{n+1}}
\eea
may be treated exactly as in the case $T<T_c$ if we replace $P$ and $Q$ by $\widehat{P}$ and $\widehat{Q}$. Thus we find 
\bea (-1)^{N+1}\widehat{D}_{N+1}
={\widehat S}_{\infty}\exp{\sum_{n=1}^{\infty}\widehat{F}^{(2n)}_{N+1}},
\eea
and hence we have
\bea D_{N}^{(+)}(t)=-{\widehat S}_{\infty}x_N^{(N)}\exp{\sum_{n=1}^{\infty}\widehat{F}^{(2n)}_{N+1}},
\eea
where we note that when $\alpha_1=0$, equation (\ref{c}) holds.





It remains to calculate $x_N^{(N)}$. We will find $x_N^{(N)}$ by iterating the procedure of section 2 of Wu \cite{W}. We define

\bea X_N(\xi)=\sum_{n=0}^{N}x_n^{(N)}\xi^n,
\eea
and thus
\bea x_N^{(N)}=\lim_{\xi \rightarrow 0}X(\xi^{-1})\xi^N
\eea
where $X_N(\xi)$ is again defined by (\ref{X}) to 
(\ref{U}) with $P(\xi)$ and $Q(\xi)$ replaced by $\widehat{P}(\xi)$ 
and $\widehat{Q}(\xi)$. For convenience we rewrite (\ref{V}), 
replacing $\xi$ with $\xi^{-1}$ as
\bea V_N(\xi)&&=-\widehat{P}(\xi)\left\{[\widehat{Q}(\xi)]_+'
+[\widehat{Q}(\xi)U_N(\xi^{-1})\xi^{-N}]_+'\right\}\nn
&&=\widehat{P}(\xi)-1-\widehat{P}(\xi)[\widehat{Q}(\xi)U_N(\xi^{-1})\xi^{-N}]_+'.
\label{Vp}
\eea
To obtain the first approximation $x^{(N)(1)}_N$ 
we replace $U(\xi)$ by 0 in (\ref{Vp}), and write

\bea V_N^{(1)}(\xi)=\widehat{P}(\xi)-1.
\label{V1a}
\eea
We use this in (\ref{Xm}) to give

\bea X_N^{(1)}(\xi^{-1})\xi^N=\widehat{Q}(\xi)[\widehat{P}(\xi^{-1})\widehat{P}(\xi)\xi^N]_+.
\label{X1a}
\eea
Thus letting $\xi$ approach 0 and using (\ref{X1a}) 
we obtain  the first approximation  $x_N^(N,1)$, which we denote 
as $G^{(1)}_N$:
\bea G^{(1)}_N=x_N^{(N)(1)}
=\frac{1}{2\pi i}\oint_{|\xi|=1} \widehat{P}(\xi^{-1})\widehat{P}(\xi)\xi^{N-1}d\xi.
\label{x1a}
\eea
We now compute the second approximation by using 
(\ref{V1a}) in (\ref{U}) to obtain 
\bea U^{(2)}(\xi^{-1})=-\widehat{Q}(\xi^{-1})[\widehat{P}(\xi^{-1})\widehat{P}(\xi)\xi^N]_-.
\label{U2a}
\eea
We use (\ref{U2a}) in (\ref{Vp}) to find
\bea V^{(2)}_N(\xi)=\widehat{P}(\xi)-1
+\widehat{P}(\xi)\big[\widehat{Q}(\xi)\widehat{Q}(\xi^{-1})\xi^{-N}
[\widehat{P}(\xi^{-1})\widehat{P}(\xi)\xi^N]_-\big]_+'.
\label{V2a}
\eea
Using this in (\ref{Xm}) we obtain
\bea X^{(2)}(\xi^{-1})\xi^N
=\widehat{Q}(\xi)\bigg\{[\widehat{P}(\xi^{-1})\widehat{P}(\xi)\xi^N]_+
+\Big[\widehat{P}(\xi^{-1})\widehat{P}(\xi)\xi^N
\big[\widehat{Q}(\xi^{-1})\widehat{Q}(\xi)\xi^{-N}
[\widehat{P}(\xi^{-1})\widehat{P}(\xi)\xi^N]_-\big]_+'\Big]_+\bigg\}.
\label{X2a}
\eea
Letting $\xi=0$ in (\ref{X2a}), we see that 

\bea x_{N}^{(N)(3)}=G^{(1)}_N+G^{(3)}_N,
\eea
where
\bea G^{(3)}_N=\frac{1}{(2\pi i)^3}&&\lim_{\epsilon \rightarrow 0}\oint_{|z_1|=1} dz_1~z_1^N\widehat{P}(z_1)\widehat{P}(z_1^{-1})
\oint_{|z_2|=1-\epsilon} dz_2~\frac{z_2^{N+1}}{1-z_1z_2}\widehat{Q}(z_2)\widehat{Q}(z_2^{-1})\nn
&&\oint_{|z_3|=1} dz_3~\frac{z_3^{N}}{1-z_2z_3}\widehat{P}(z_3)\widehat{P}(z_3^{-1}).
\eea
Continuing in the same way we may find
\bea x_{N}^{(N)(2n+1)}=\sum_{k=0}^{n}G^{(2k+1)}_N,
\eea
and thus 
\bea D_N^{(+)}(t)=-(1-t)^{1/4}\sum_{n=0}^{\infty}G_N^{(2n+1)}\exp{\sum_{m=1}^{\infty}\widehat{F}_{N+1}^{(2m)}}
\eea
where $\widehat{F}_N^{(2n)}$ is defined in (\ref{Fh}) and $G_N^{(2n+1)}$ is defined in (\ref{G}).

If we note that the $G^{(2n+1)}_N$ is the negative of the
$G^{(2n+1)}_N$ of \cite{M} and set $\alpha_1=0$ we have proven (6) 
of \cite{M} with $G^{(2n+1)}_N$ given by (34) of \cite{M}. 
 
\section{The form factor expansion for $T<T_c$}
\label{below2}
We have showed in section \ref{below} that the correlation function 
$D_{N}^{(-)}$ can be written in an exponential form given by 
(\ref{effei}) and (\ref{F}). In this section we will show that 
$D_{N}^{(-)}$ can be written as a form factor expansion given by 
equations (\ref{ffm}) and (\ref{f}).

We wish to rewrite (\ref{effei}) as a form factor expansion and use an argument similar to that made by Nappi \cite{Na} to find the functions $f_N^{(2n)}$. To do this, we denote by a partition $\pi$ of the number $n$ a set of pairs $\pi=\{(n_i,m_i)\}_{i=1}^{\nu(\pi)}$ such that $n_i\neq n_j$ if $i\neq j$ and
\bea \sum_{i=1}^{\nu(\pi)}n_im_i=n.
\label{partn}
\eea
We define $\mathcal{P}(n)$ to be the set of all such partitions. For instance, the partitions of the number 3 are


\begin{eqnarray}
3 = \left\lbrace \begin{array}{ll}
           1\cdot 3,~\nu=1 \\
           3\cdot 1,~\nu=1\\
           1\cdot 1+2\cdot 1,~\nu=2 &\mbox{}. 
\end{array} \right.
\label{three}
\end{eqnarray}
Thus the exponential of (\ref{effei}) may be expanded, and we find
\bea f_N^{(2n)}=\sum_{\pi \in \mathcal{P}(n)}\prod_{i=1}^{\nu(\pi)}\frac{1}{m_i!}\left(F_N^{(2n_i)}\right)^{m_i},
\label{resn}
\eea
where the sum is over the set of partitions $\mathcal{P}(n)$ of the number $n$. Thus $f_N^{(2n)}$ is the sum of all $2n$ dimensional integrals in (\ref{effei}). Explicitly 
\bea f_N^{(2n)}=&&\frac{(-1)^n}{(2\pi)^{2n}}\lim_{\epsilon \rightarrow 0}\prod_{i=1}^{2n}\oint_{|z_i|=1-\epsilon} dz_i~z_{i}^N\prod_{j=1}^{n} Q(z_{2j-1})Q(z_{2j-1}^{-1})P(z_{2j})P(z_{2j}^{-1})\frac{1}{1-z_{2j-1}z_{2j}}\nn
&&\sum_{\pi \in \mathcal{P}(n)} \prod_{k=1}^{\nu(\pi)}(-1)^{m_k}\frac{1}{m_k!n_k^{m_k}}\prod_{p=1}^{m_k}\prod_{q=1}^{n_k}\frac{1}{1-z_{\sum_{r=1}^{k-1}2m_rn_r+2(p-1)n_k+2q}z_{\sum_{r=1}^{k-1}2m_rn_r+2(p-1)n_k+(2q\oplus _{\pi,k}1)}}
\label{f2n1n}
\eea
where
\begin{eqnarray}
2q\oplus _{\pi,k}1 := \left\lbrace \begin{array}{ll}
            2q+1  & \mbox{if $q<n_k$}\\
           1  & \mbox{if $q=n_k$}.
\end{array} \right.
\label{newopn}
\end{eqnarray}
We see that a partition $\pi$ divides the integrand into $\sum_{k=1}^{\nu(\pi)}m_k$ loops, and that there are $m_k$ loops of length $n_k$. As an illustration,
\bea f_N^{(6)}=&&-\frac{1}{(2\pi)^{6}}\lim_{\epsilon \rightarrow 0}\prod_{i=1}^{6}\oint_{|z_i|=1-\epsilon} dz_i\prod_{j=1}^{3} Q(z_{2j-1})Q(z_{2j-1}^{-1})P(z_{2j})P(z_{2j}^{-1}) \prod_{l=1}^{6}z_{l}^N\nn
&&\frac{1}{1-z_1z_2}\frac{1}{1-z_3z_4}\frac{1}{1-z_5z_6}\nn
&&\bigg(-\frac{1}{3!}\frac{1}{1-z_2z_1}\frac{1}{1-z_4z_3}\frac{1}{1-z_6z_5}-\frac{1}{3}\frac{1}{1-z_2z_3}\frac{1}{1-z_4z_5}\frac{1}{1-z_6z_1}\nn
&&+\frac{1}{2}\frac{1}{1-z_2z_1}\frac{1}{1-z_4z_5}\frac{1}{1-z_6z_3}\bigg)
\label{f61n}
\eea
The first term in the bracket of the right hand side of (\ref{f61n}) comes from $\pi_1=\{(1,3)\}$, the second from $\pi_2=\{(3,1)\}$ and the third from $\pi_3=\{(1,1),~(2,1)\}$.
We would like to show that 
\bea f_N^{(2n)}=&&\frac{1}{n!(2\pi)^{2n}}\lim_{\epsilon \rightarrow 0}\prod_{i=1}^{2n}\oint_{|z_i|=1-\epsilon}dz_i ~z_{i}^{N}
\prod_{j=1}^{n}Q(z_{2j-1})Q(z_{2j-1}^{-1})P(z_{2j})P(z_{2j}^{-1})\frac{1}{1-z_{2j-1}z_{2j}}\nn
&&\sum_{\sigma \in S_n} \textrm{sign}(\sigma)\prod_{k=1}^{n}\frac{1}{1-z_{2k}z_{\sigma(2k-1)}},
\label{spsn}
\eea
where $S_n$ is the group of permutations of the $n$ elements $\{2i-1\}_{i=1}^n$. For instance,
\bea S_3=\{(1)(3)(5),~(13)(5),~(15)(3),~(35)(1),~(135),~(153)\}
\label{S3}
\eea
where the loop $(abc)$ means the permutation $a\rightarrow b\rightarrow c\rightarrow a$. We say that two permutations $\sigma_1$ and $\sigma_2$ in $S_n$ are equivalent if for every loop in $\sigma_1$ there is one and only one loop of equal length in $\sigma_2$. Then $\sigma_1$ and $\sigma_2$ will also have the same signature. We write the equivalence class of an element $\sigma$ as $[\sigma]$. We denote by $E_n$ the set of equivalence classes of $S_n$. As an example, we have 
\bea E_3=\{[(1)(3)(5)],~[(13)(5)],~[(135)]\} 
\label{E3n}
\eea
We will show that there is a bijection between $\mathcal{P}(n)$ and $E_n$. It is clear that $|\mathcal{P}(3)|=|E_3|=3$. We will now prove the general case.

We will now calculate $|[\sigma]|$, the number of elements of the equivalence class of a permutation $\sigma$. 
We consider some $\sigma \in S_n$, and construct $[\sigma]$ as follows. We choose freely from $n$ elements, and divide them into $\sum_{i=1}^{\nu}m_i$ loops such that there are $m_i$ loops with $n_i$ elements, without distinguishing between loops with the same number of elements. There are
\bea \frac{n!}{\prod_{i=1}^{\nu}(n_i!)^{m_i}m_i!}
\label{waysn}
\eea
ways of doing this. There are $(n_i-1)!$ ways of ordering a loop of $n_i$ elements. Hence, the there are 
\bea |[\sigma]|=\frac{n!\prod_{i=1}^{\nu}((n_i-1)!)^{m_i}}{\prod_{j=1}^{\nu}(n_j!)^{m_j}m_j!}=\frac{n!}{\prod_{i=1}^{\nu}n_i^{m_i}m_i!}
\label{ways2n}
\eea
ways of choosing the elements. 
The signature of any element of the equivalence class $[\sigma]$ corresponding to $\pi$ is 
\be \textrm{sign}(\sigma)=(-1)^n\prod_{k=1}^{\nu(\pi)}(-1)^{m_k}. 
\ee
If we identify every equivalence class with one of its representatives, then it follows from (\ref{f2n1n}) that
\bea f_N^{(2n)}=&&\frac{1}{n!(2\pi)^{2n}}\lim_{\epsilon \rightarrow 0}\prod_{i=1}^{2n}\oint_{|z_i|=1-\epsilon} dz_i~z_{i}^N\prod_{j=1}^{n} Q(z_{2j-1})Q(z_{2j-1}^{-1})P(z_{2j})P(z_{2j}^{-1})~\frac{1}{1-z_{2j-1}z_{2j}} \nn
&&\sum_{\sigma \in E_n} \textrm{sign}(\sigma)~|[\sigma]|\prod_{k=1}^n\frac{1}{1-z_{2k}z_{\sigma(2k-1)}}.
\label{f2n2n}
\eea
Now (\ref{spsn}) follows. By symmetry of the odd variables, (\ref{spsn}) can be rewritten as 
\bea f_N^{(2n)}=&&\frac{1}{(n!)^2(2\pi)^{2n}}\lim_{\epsilon \rightarrow 0}\prod_{i=1}^{2n}\oint_{|z_i|=1-\epsilon}dz_i~z_{i}^{N}
\prod_{j=1}^{n}Q(z_{2j-1})Q(z_{2j-1}^{-1})P(z_{2j})P(z_{2j}^{-1})\nn
&&\left(\sum_{\sigma \in S_n} \textrm{sign}(\sigma)\prod_{k=1}^{n}\frac{1}{1-z_{2k}z_{\sigma(2k-1)}}\right)^2.
\label{sppn}
\eea
Finally we note that the factor of the integrand of (\ref{sppn}) 
\be \sum_{\sigma \in S_n} \textrm{sign}(\sigma)\prod_{k=1}^{n}\frac{1}{1-z_{2k}z_{\sigma(2k-1)}}
\ee
is zero if for any $i \neq j$, $z_{2i}=z_{2j}$ or $z_{2i-1}=z_{2j-1}$. Therefore 
\bea \sum_{\sigma \in S_n} \textrm{sign}(\sigma)\prod_{k=1}^{n}\frac{1}{1-z_{2k}z_{\sigma(2k-1)}}
=A_n\prod_{k=1}^n\prod_{l=1}^n\frac{1}{(1-z_{2k-1}z_{2l})}
\prod_{1\leq p<q\leq n}(z_{2p-1}-z_{2q-1})(z_{2p}-z_{2q}).
\eea
By letting $z_{2n}=z_{2n-1}=0$ we find that
\be A_n=A_{n-1}.
\ee
Since $A_1=1$, it follows that $A_n=1$ for all $n$. Hence
we obtain the desired result (\ref{f}).

\section{The form factor expansion for $T>T_c$}
\label{above2}
Above $T_c$, $D_{N}^{(+)}$ has a form factor expansion given by (\ref{rffeai}),
where
\bea f_N^{(2n+1)}=\sum_{k=0}^{n}G_N^{(2k+1)}\widehat{f}_{N+1}^{(2n-2k)}
\label{faboven}
\eea 
and $\widehat{f}_N^{(2n)}$ is given by (\ref{spsn}) but 
with $P$ and $Q$ replaced by $\widehat{P}$ and $\widehat{Q}$. 
$G^{(2n+1)}_N$ is given by (\ref{G}).
Hence it follows from (\ref{spsn}), (\ref{faboven}) and (\ref{G}) that 
\bea f_N^{(2n+1)}=&&-\frac{i}{(2\pi)^{2n+1}}\lim_{\epsilon \rightarrow 0}\prod_{i=1}^{2n+1}\oint_{|z_i|=1-\epsilon}dz_i ~z_i^{N+1}\prod_{l=1}^{n+1}\widehat{P}(z_{2l-1})\widehat{P}(z_{2l-1}^{-1})\prod_{m=1}^{n}\widehat{Q}(z_{2m})\widehat{Q}(z_{2m}^{-1})\frac{1}{z_{2n+1}}\prod_{p=1}^{n}\frac{1}{1-z_{2p-1}z_{2p}}\nn
&&\sum_{k=0}^{n}(-1)^k\frac{1}{(n-k)!}\frac{1}{z_{2n-2k+1}}\sum_{\sigma \in S_{n-k}}\textrm{sign}(\sigma)\prod_{q=1}^{n-k}\frac{1}{1-z_{2q-1}z_{\sigma(2q)}}\prod_{s=n-k+1}^{n}\frac{1}{1-z_{2s}z_{2s+1}}.
\label{fa2n}
\eea
As an example,
\bea f_N^{(5)}=&&-\frac{i}{(2\pi)^{5}}\lim_{\epsilon \rightarrow 0}\prod_{i=1}^{5}\oint_{|z_i|=1-\epsilon}dz_i ~z_i^{N+1}\prod_{l=1}^{3}\widehat{P}(z_{2l-1})\widehat{P}(z_{2l-1}^{-1})\prod_{m=1}^{2}\widehat{Q}(z_{2m})\widehat{Q}(z_{2m}^{-1})\frac{1}{z_{5}}\frac{1}{1-z_{1}z_{2}}\frac{1}{1-z_{3}z_{4}}\nn
&&\left(\frac{1}{2}\frac{1}{z_{5}}\left(\frac{1}{1-z_{1}z_{2}}\frac{1}{1-z_{3}z_{4}}-\frac{1}{1-z_{1}z_{4}}\frac{1}{1-z_{2}z_{3}}\right)-\frac{1}{z_{3}}\frac{1}{1-z_{1}z_{2}}\frac{1}{1-z_{4}z_{5}}+\frac{1}{z_{1}}\frac{1}{1-z_{2}z_{3}}\frac{1}{1-z_{4}z_{5}}\right).
\label{ff2n}
\eea
Let $(i_1^{(k)},...,i_n^{(k)}):=(1,...,n-k,n-k+2,...,n+1)$. It follows by symmetry that (\ref{fa2n}) can be rewritten as
\bea f_N^{(2n+1)}=&&-\frac{i}{(2\pi)^{2n+1}}\lim_{\epsilon \rightarrow 0}\prod_{i=1}^{2n+1}\oint_{|z_i|=1-\epsilon}dz_i ~z_i^{N+1}\prod_{l=1}^{n+1}\widehat{P}(z_{2l-1})\widehat{P}(z_{2l-1}^{-1})\prod_{m=1}^{n}\widehat{Q}(z_{2m})\widehat{Q}(z_{2m}^{-1})\nn
&&\frac{1}{n+1}\sum_{r=0}^{n}(-1)^r\frac{1}{z_{2n-2r+1}}\prod_{p=1}^{n}\frac{1}{1-z_{2i_p^{(r)}-1}z_{2i_p^{(r)}}}\nn
&&\frac{1}{n!}\sum_{k=0}^{n}(-1)^k\frac{1}{z_{2n-2k+1}}\sum_{\sigma \in S_{n}}\textrm{sign}(\sigma)\prod_{q=1}^{n}\frac{1}{1-z_{2i_q^{(k)}-1}z_{\sigma(2i_q^{(k)})}}.
\label{fa3n}
\eea
In particular
\bea f_N^{(5)}=&&-\frac{i}{(2\pi)^{5}}\lim_{\epsilon \rightarrow 0}\prod_{i=1}^{5}\oint_{|z_i|=1-\epsilon}dz_i ~z_i^{N+1}\prod_{l=1}^{3}\widehat{P}(z_{2l-1})\widehat{P}(z_{2l-1}^{-1})\prod_{m=1}^{2}\widehat{Q}(z_{2m})\widehat{Q}(z_{2m}^{-1})\nn
&&\frac{1}{3}\left(\frac{1}{z_{5}}\frac{1}{1-z_{1}z_{2}}\frac{1}{1-z_{3}z_{4}}-\frac{1}{z_{3}}\frac{1}{1-z_{1}z_{2}}\frac{1}{1-z_{4}z_{5}}+\frac{1}{z_{1}}\frac{1}{1-z_{2}z_{3}}\frac{1}{1-z_{4}z_{5}}\right)\nn
&&\bigg\{\frac{1}{2}\frac{1}{z_{5}}\left(\frac{1}{1-z_{1}z_{2}}\frac{1}{1-z_{3}z_{4}}-\frac{1}{1-z_{1}z_{4}}\frac{1}{1-z_{2}z_{3}}\right)\nn
&&-\frac{1}{2}\frac{1}{z_{3}}\left(\frac{1}{1-z_{1}z_{2}}\frac{1}{1-z_{4}z_{5}}-\frac{1}{1-z_{1}z_{4}}\frac{1}{1-z_{2}z_{5}}\right)\nn
&&+\frac{1}{2}\frac{1}{z_{1}}\left(\frac{1}{1-z_{2}z_{3}}\frac{1}{1-z_{4}z_{5}}-\frac{1}{1-z_{3}z_{4}}\frac{1}{1-z_{2}z_{5}}\right)\bigg\}.
\label{ff3n}
\eea
Since all permutations of the even elements are present in the sum $\sum_{k=0}^n$, symmetry allows the permutation of all even elements in the sum $\sum_{r=0}^n$. But the sum $\sum_{k=0}^n\sum_{\sigma \in S_n}$ may be rewritten as the sum $\sum_{\sigma \in S_{n+1}}$ of permutations of the odd elements. Therefore
\bea f_N^{(2n+1)}=-\frac{i}{(2\pi)^{2n+1}}\lim_{\epsilon \rightarrow 0}\prod_{i=1}^{2n+1}\oint_{|z_i|=1-\epsilon}dz_i ~z_i^{N+1}\prod_{l=1}^{n+1}\widehat{P}(z_l)\widehat{P}(z_l^{-1})\prod_{m=1}^{n}\widehat{Q}(z_m)\widehat{Q}(z_m^{-1})\nn
\frac{1}{n!(n+1)!}\left(\sum_{\sigma \in S_{n+1}}\textrm{sign}(\sigma)\frac{1}{z_{\sigma(2n+1)}}\prod_{q=1}^{n}\frac{1}{1-z_{\sigma(2q-1)}z_{2q}}\right)^2.
\label{fa5n}
\eea
An argument similar to the one given in section \ref{below2} shows that 
\bea \sum_{\sigma \in S_{n+1}}\textrm{sign}(\sigma)\frac{1}{z_{\sigma(2n+1)}}\prod_{q=1}^{n}\frac{1}{1-z_{\sigma(2q-1)}z_{2q}}=&&\prod_{j=1}^{n+1}\frac{1}{z_{2j-1}}\prod_{k=1}^{n}\frac{1}{1-z_{2j-1}z_{2k}}\nn
&&\prod_{1\leq l<m\leq n+1}(z_{2l-1}-z_{2m-1})\prod_{1\leq p<q\leq n}(z_{2p}-z_{2q}).
\label{sqn}
\eea
Thus $f^{(2n+1)}_N$ is given by (\ref{fa6n}) as desired.

\section{Discussion}
The exponential and the form factor representations derived in this
paper for $\langle \sigma_{0,0}\sigma_{0,N}\rangle$ and
$\langle \sigma_{0,0}\sigma_{N,N}\rangle$ are considerably simpler
that the corresponding representations which may be found in 
\cite{M2}-\cite{Ni2}. The representations of this paper must of course
be equal to the corresponding results of \cite{M2}-\cite{Ni2} but as
mentioned in the introduction even the equality of the form of $F^{(2)}_N$
found by Wu \cite{W} with the form found by Cheng and Wu \cite{CW} has
not been directly demonstrated in the literature. The form factor
representations for $\langle \sigma_{0,0}\sigma_{N,N}\rangle$
proven here are in close correspondence with formulas given by Jimbo
and Miwa \cite{JM} in their proof of the Painleve VI representation of
the diagonal Ising correlations. The connection which the form factor
representations of this paper have with the PVI equation of \cite{JM}
have been extensively investigated in \cite{M}. However, 
the representations of this paper are valid also for  
$\langle \sigma_{0,0}\sigma_{0,N}\rangle$ and, as noted in the introduction,
for much more general case which suggests that there are
generalizations of \cite{JM} which have not yet been uncovered. In
particular the relation of $\langle \sigma_{0,0}\sigma_{0,N}\rangle$
to isomonodromic deformation theory remains to be investigated.

\label{dis}

{\center{\bf Acknowledgments}}

We would like to thank Prof. J.-M. Maillard and Prof. N. Zenine 
for many insightful discussions and Prof. J.-M. Maillet for discussions
and hospitality at Ecole Normale Superieur Lyon 
where part of this work was performed. This work is supported in  
part by NSF grant DMR-0302758.

\end{document}